\documentclass{ws-procs975x65}
	
\begin{document}



\title{Thermal and chemical evolution of the primordial clouds in warm dark matter models 
with keV sterile neutrinos in one-zone approximation\footnote{The
work of P. L. B. and J. S. was supported by the Pierre Auger grant
05CU 5PD1/2 via the Deutsches Elektronen-Synchrotron (DESY)/
Bundesministerium fuer Bildung und Forschung (BMBF). The
work of A. K. was supported in part by the DOE grant DE-FG 03-
91ER40662 and by the NASA ATP grants NAG 05-10842 and
NAG 05-13399.}}

\author{Jaroslaw Stasielak\footnote{J. S. thanks the organizers of Marcel Grossmann
meeting for financial support.}}

\address{Institute of Physics, Jagiellonian University, Krak\'ow, 30-059, Poland, and\\
Max-Planck Institute for Radioastronomy, Bonn, D-53121, Germany, and\\
Department of Physics and Astronomy, University of Bonn, D-53121, Germany\\
\email{stasiela@th.if.uj.edu.pl}}

\author{Peter L. Biermann}

\address{Max-Planck Institute for Radioastronomy, Bonn, D-53121, Germany, and\\
Department of Physics and Astronomy, University of Bonn, D-53121, Germany, and\\
Department of Physics and Astronomy, University of Alabama, AL 35487, Tuscaloosa, USA\\
\email{pbiermann@mpifr-bonn.mpg.de}}

\author{Alexander Kusenko\footnote{A. K. thanks the CERN Theory unit and Ecole
Polytechnique Fe´de´rale de Lausanne (EPFL) for hospitality during
his visit.}}
\address{Department of Physics and Astronomy, University of California, CA 90095-1547, Los
Angeles, USA\\
\email{kusenko@ucla.edu}}


\begin{abstract}

We follow the evolution of the baryonic top-hat overdensity in a single-zone approximation. Our goal is to juxtapose
the evolution of the gas temperature in the primordial clouds in the lambda cold dark matter model and the warm dark matter model with keV sterile neutrinos and to check the effects of their decays, into one X-ray photon and one active neutrino, on the structure formation. We find that, in all the cases
we have examined, the overall effect of sterile dark matter is to facilitate
the cooling of gas and to reduce the minimal mass of the halo able to collapse. Hence, we conclude that X-rays from the decays of dark matter in the form of
sterile neutrinos can help the early collapse of gas clouds and the
subsequent star formation.
\end{abstract}

\bodymatter

\vline

 Recent work has showed several inconsistencies between the predictions of N-body simulations of collisionless cold dark matter (CDM) and the observations \cite{numerical}. Perhaps, a better understanding of CDM on small scales will resolve these discrepancies. It is true, however, that all these problems altogether can be solved by suppression of the primordial power spectrum of scalar density fluctuations on small scales. This can be done in warm dark matter (WDM) models via the non-negligible kinetic energy of the dark matter particles.

One attractive candidate for WDM is sterile neutrino (SN) with mass of several keV and a small mixing angle with the ordinary neutrino. Such a particle would be a natural part of minimal extension of the standard model ($\nu$MSM) in which existence of three SNs would be able to explain the masses of active neutrinos\cite{mass}, the baryon asymmetry of the universe\cite{asymmetry} and the abundance of dark matter\cite{abundance}. Furthermore, SN with mass in the keV range can simultaneously solve another seemingly unrelated astrophysical puzzle, namely, the origin of the rapid motions of pulsars\cite{kick}. It can also bring the supernova calculation in better agreement with the observation\cite{supernova} and help the formation of super-massive black holes in the early universe\cite{hole}.

\begin{figure}[t]
\psfig{file=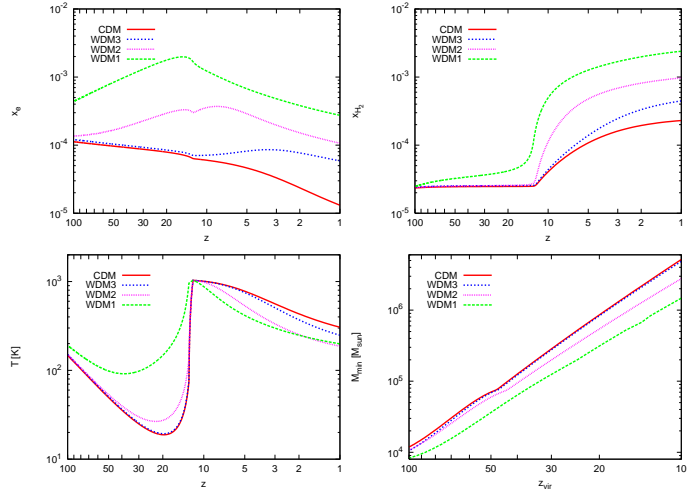,width=4in,bbllx=0, bblly=50, bburx=540, bbury=400,clip=}
\caption{{\it Top left}, {\it top right} and {\it bottom left}: Evolution of ionization fraction, $H_{2}$ fraction and temperature with redshift for different models. In each case, the mass of the primordial cloud is equal to $M=10^6 M_{\odot}$ and virialization redshift to $z_{vir}=12$. 
{\it Bottom right}: Dependence of the minimal mass of primordial halo able to collapse on its virialization redshift. Models we have used in calculation are following: $m_{s}=25$ keV and $\sin^2 \theta=3 \times 10^{-12}$ (WDM1), $m_{s}=15$ keV and $\sin^2 \theta=3 \times 10^{-12}$ (WDM2), $m_{s}=3.3$ keV and $\sin^2 \theta=3 \times 10^{-9}$ (WDM3), and CDM.} 
\label{sbk:fig1}
\end{figure}

Although dark-matter SNs are stable on cosmological time scales, they nevertheless decay. The most prominent "visible" mode is decay into one active neutrino and one photon, $\nu_s \rightarrow \nu_a \gamma$. It produces an X-ray background radiation that has a two-fold effect on the collapsing clouds of hydrogen in the early universe.  First, the X-rays ionize gas and cause an increase in the fraction of molecular hydrogen\cite{bier}, which makes it easier for the gas to cool and to form stars.  Second, the same X-rays deposit a certain amount of heat, which could thwart the cooling of gas.
Moreover, it is well known that suppression on small scales in WDM models delays structure formation and can lead, in principle, to inconsistency with WMAP measurements\cite{yoshida}. It is, therefore, important to check cosmological implication of keV SNs via examination of their decays on star formation and re-ionization of the universe.

We follow the evolution of the baryonic top-hat overdensity in a single-zone approximation \cite{tegmark,stas}, assuming that the density of the collapsing cloud is constant after virialization. In order to take into account the effects of SNs decays inside the collapsing halo, we have solved radiative transfer equation for the spherically symmetric clouds with the uniform density\cite{transfer}. In addition, we have included absorption of the X-rays from the SNs decays by both $H$ and $He$. These are improvements with respect to \citen{stas}, where one can find the details of our code. 

\Fref{sbk:fig1} shows our results.
We can clearly see that the overall effect of SN decay is to enhance ionization fraction, $H_{2}$ fraction and to speed up the cooling of the gas in the primordial halos. The minimal mass of the clouds able to collapse is reduced in all WDM models we have used.
It is worth to note that inclusion of SN free-streaming length can change substantially these results and lead to the delay of structure formation with respect to the CDM model\cite{ripamonti}.
However, if one compares two WDM models, namely the one with SNs which cannot decay and the second one with decaying SNs, then the gas cooling should be enhanced in the latter case. If this is true then our conclusion will not change. Our work shows the importance of the dark matter decays in the structure formation in the early universe.

We thank G. Gilmore, M. Mapelli, E. Ripamonti and M. Shaposhnikov for very helpful discussions and comments.  

\vfill

\end{document}